\documentclass{aastex}
\usepackage{epsfig}
\usepackage{emulateapj5}

\newcommand{\feiii}{\ion{Fe}{3}}
\newcommand{\nii}{\ion{N}{2}}
\newcommand{\nI}{\ion{N}{1}}
\newcommand{\hh}{H$_{2}$}
\newcommand{\kms}{km~s$^{-1}$}
\newcommand{\cm}{cm$^{-2}$}
\newcommand{\fuse}{{\it FUSE}}
\newcommand{\oph}{51~Oph}
\newcommand{\bp}{$\beta$~Pictoris}
\newcommand{\lb}{$\lambda$}
\newcommand{\om}{\ion{O}{1}~($^{1}D$)}
\newcommand{\nd}{\ion{N}{1}~($^{2}D$)}
\newcommand{\ndf}{\ion{N}{1}~($^{2}D_{5/2}$)}
\newcommand{\ndt}{\ion{N}{1}~($^{2}D_{3/2}$)}
\newcommand{\np}{\ion{N}{1}~($^{2}P$)}
\newcommand{\sd}{\ion{S}{2}~($^{2}D$)}
\newcommand{\sdt}{\ion{S}{2}~($^{2}D_{3/2}$)}
\newcommand{\sdf}{\ion{S}{2}~($^{2}D_{5/2}$)}

\shorttitle{Infalling planetesimals in the 51~Oph CS disk}
\shortauthors{Roberge et al.}

\begin{document}

\title{{\it FUSE} observations of possible infalling planetesimals in the 
51~Ophiuchi circumstellar disk}

\author{A.~Roberge\altaffilmark{1}, P.~D.~Feldman\altaffilmark{1}, 
A.~Lecavelier des Etangs\altaffilmark{2},
A.~Vidal-Madjar\altaffilmark{2}, M.~Deleuil\altaffilmark{3}, 
J.-C.~Bouret\altaffilmark{3}, R.~Ferlet\altaffilmark{2} and
H.~W.~Moos\altaffilmark{1}} 

\email{akir@pha.jhu.edu}

\altaffiltext{1}{Department of Physics and Astronomy, 
Johns Hopkins University, 3400 N. Charles St., Baltimore, MD 21218}

\altaffiltext{2}{Institut d'Astrophysique de Paris, CNRS, 98bis Bd Arago, 
F-75014 Paris, France}

\altaffiltext{3}{Laboratoire d'Astrophysique de Marseille, BP 8, F-13376 
Marseille Cedex 12, France}

\begin{abstract}

We present the first observations of the circumstellar (CS) disk system 
51~Ophiuchi with the {\it Far Ultraviolet Spectroscopic Explorer}. 
We detect several absorption lines arising from the 
unusual metastable atomic species \nd, \np, and \sd. These 
levels lie 1.8 -- 3.6 eV above the ground level and have radiative decay 
lifetimes of 2 days or less, indicating that the lines arise 
from warm CS gas. 
The high S/N \fuse\ spectra, obtained six days apart, also show 
time-variable absorption features arising from \nI, \nii, \om, and \feiii,
which are redshifted with respect to the stellar velocity.  
The resolved redshifted absorption extends over many tens of \kms\ 
(40 for \nI, 100 for \nii, 65 for \om, and 84 for \feiii).
We calculate column densities for all the variable infalling CS 
gasses, using the apparent optical depth method. The \feiii\ and
\nii\ infalling gasses must be produced through collisional
ionization, and the ionization fraction of nitrogen suggests 
a gas temperature between 20000 and 34000 K. The infalling
gas shows a peculiar, non-solar composition, with nitrogen and iron 
more abundant than carbon.  We also set upper limits on the 
line-of-sight column densities of H$_2$ and CO. 
These observations strengthen the connection between 51~Oph and 
the older debris-disk system $\beta$~Pictoris, and indicate that 
there may be infalling planetesimals in the 51~Oph system.

\end{abstract}


\section{Introduction}\label{sec:intro}

\objectname[HD 158643]{51~Ophiuchi} (HD~158643) is a young 
B9.5 Ve star ($\sim 3 \times 10^{5}$ yrs old; \citet{van98}), 
showing a large infrared emission excess arising 
from circumstellar (CS) dust. This dust is presumed to be located 
in a disk similar to that observed around $\beta$~Pictoris \citep{wat88}.  
However, although 51~Oph lies only 
$131^{+17}_{-13}$ pc away from the Sun \citep{van98}, its CS dust disk 
has not yet been imaged. {\it ISO} spectroscopy of the 51~Oph CS 
dust shows the presence of crystalline silicates, which are seen 
in the spectra of $\beta$~Pic and Solar System meteorites and comets, 
but not in spectra of interstellar (IS) dust \citep{wael96}. The disk of 51~Oph
is probably close to edge-on, because the very large projected rotational
velocity of the star ($v \ {\rm sin} i = 270$ \kms; \citet{dunk97}) 
suggests that our line-of-sight is parallel to the equatorial plane
of the system. 

51~Oph also has a CS gas disk, as evidenced by double-peaked hydrogen
emission \citep{slett82} and narrow, non-IS absorption features 
in the ultra-high-resolution spectrum of the star \citep{craw97}. 
Cold CS \ion{C}{1} gas ($T \approx 20$ K) has also been observed 
toward 51~Oph; this species has a very short 
photoionization lifetime and must be continuously replenished 
\citep{lecav97a}. Most interestingly, {\it IUE} spectra
of 51~Oph show variable infalling gas absorption features arising 
from species with a wide range of ionization states, including
\ion{C}{4} and \ion{Si}{4}, which must be produced by collisional
ionization \citep{gra93}. Such variable features are an  
indication of the presence of infalling planetesimals
in the 51~Oph system (for a review of this phenomenon in the 
well-studied $\beta$~Pictoris system, see \citet{vid98}). 
All this evidence together strongly suggests that 51~Oph is 
similar to $\beta$~Pictoris, and represents a young planetary 
system with a debris disk of gas and dust. 

In this paper, we present the first {\it Far Ultraviolet
Spectroscopic Explorer} (\fuse) spectra of 51~Oph.
We observe several narrow CS absorption lines arising from the unusual
metastable species \nd, \np, and \sd, and 
calculate the redshifts and equivalent widths of these lines.
We also set low upper limits on the column densities of \hh\ and CO 
toward the star, which has interesting consequences 
in light of the recent detection of large quantities of warm molecular 
gasses in {\it ISO} spectra of 51~Oph \citep{van2001}.
The high S/N \fuse\ spectra were taken 6 days apart, permitting us to 
detect time-variable redshifted absorption from four species, \feiii, \nii, 
\ion{N}{1}, and metastable \om. We examine the velocity 
structure of the resolved variable features, and carefully determine 
column densities of the infalling gasses. 
In \S\ref{sec:dis}, we investigate the unusual 
composition of the infalling gas, and demonstrate that
it is most likely produced from infalling planetesimals, as in the
case of $\beta$~Pictoris. 

\section{Observations and Data Reduction}\label{sec:obs}

\begin{figure}
\begin{center}
\hspace*{-0.2in}
\epsfig{file=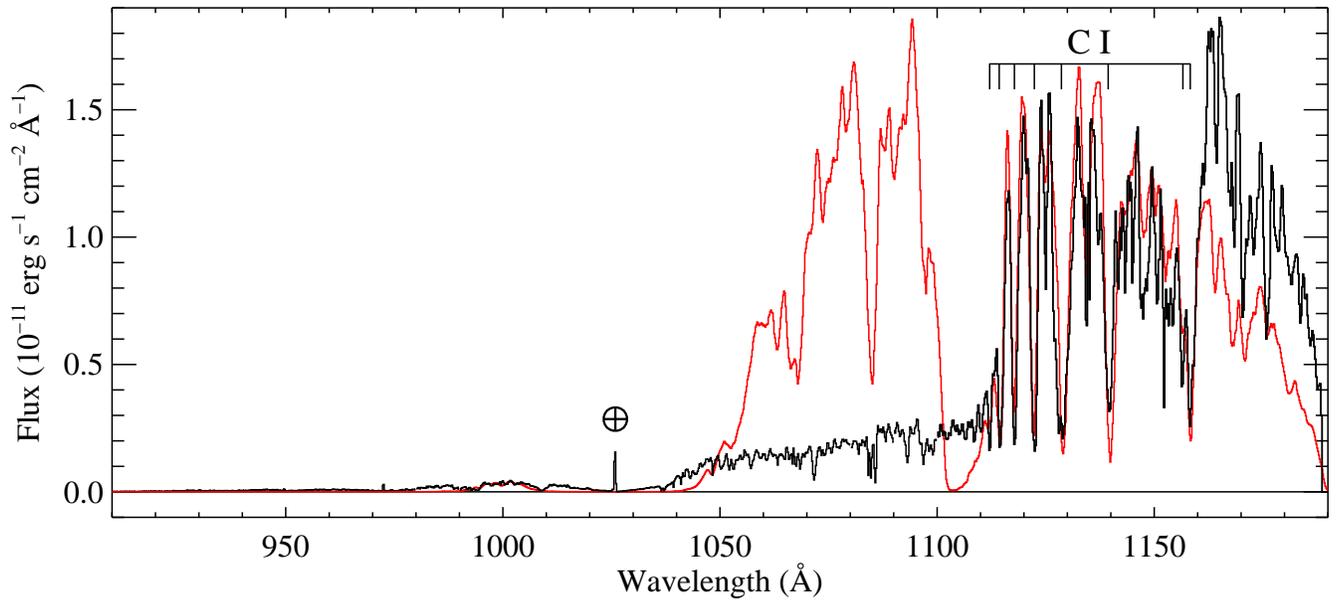}
\caption{Overview of the entire \oph\ \fuse\ spectrum.\label{fig:over}
The data from the two days of observation have been averaged together and
rebinned by a factor of 32 for this plot. 
The synthetic stellar spectrum described in the text is shown in red.
The prominent \ion{C}{1} photospheric absorption lines between 
1110 \AA\ and 1160 \AA\ are indicated. The Lyman-$\beta$ airglow line 
is marked with~$\oplus$.}
\end{center}
\end{figure}

51~Oph was observed with {\it FUSE} on 2000 August 29 and  
2000 September 4. The complete spectrum appears in 
Figure~\ref{fig:over}. The data were obtained using the low-resolution 
$30 \arcsec \times 30 \arcsec$ aperture, covering the wavelength range 
905 -- 1187~\AA. The \fuse\ satellite has four coaligned optical
channels (LiF~1, SiC~1, LiF~2, SiC~2), forming two nearly identical 
``sides'', each consisting of a LiF grating, a SiC grating, and a detector.
Each detector is divided into two independent segments (a and b),
separated by a small gap. The entire wavelength range is therefore
covered with eight partially overlapping spectra that fall on different
portions of the two detectors (LiF~1a, LiF~1b, etc.).  A discussion of the
on-orbit performance of \fuse\ may be found in \citet{sah00}. 
The data were calibrated with the CALFUSE 1.8.7 pipeline processing 
software. Detailed information on this software is available at 
\url{http://fuse.pha.jhu.edu/analysis/pipeline\_reference.html}. 
Since none of the observed lines were intrinsically narrow enough 
to use them to determine the true spectral resolution of our data, 
we conservatively assume 20~\kms\ spectral resolution 
($R~=~\lambda/\Delta\lambda~=~15000$).

The total exposure time of the first observation was 
9974~s, divided into 23 exposures, and the total exposure time of the second
9523~s, divided into 20 exposures. Since the target was not
exactly centered within the aperture in every exposure, the resulting
wavelength shifts were corrected by aligning the individual exposures
before co-addition, using a linear cross-correlation procedure. 
Therefore, the image of the star usually fell on a slightly 
different portion of the detector in each exposure, greatly reducing the 
fixed pattern noise in the final co-added spectra. In each observation, 
we obtained the following signal-to-noise ratios per 
resolution element: 69 in LiF~1b near 1130~\AA, 20 in SiC~2b near 
1080~\AA, and 36 in LiF~1a near 1050~\AA.

The wavelength scales were corrected for the sign error in the 
CALFUSE 1.8.7 heliocentric velocity calculation. The uncertainty in the 
relative wavelength calibration of the data is  about 5 pixels 
($\sim 9$ \kms). In order to determine the absolute wavelength calibration 
of the data, we used a rotationally broadened synthetic stellar 
spectrum, overplotted in Figure~\ref{fig:over}, which included the 
photospheric \ion{C}{1} lines seen in the data \citep{chay01}. 
This synthetic spectrum was calculated at the 
\fuse\ spectral resolution, assuming solar metallicity and the 
following stellar parameters: 
$T_{eff} = 10000, \ \log \, g = 4.0, \ v \, \sin \, i = 270$ \kms.
The synthetic spectrum does not fit the data well between about 1110 \AA\
and 1040 \AA; this may be due to non-LTE effects.  
A paper on improved modeling of the FUV spectra of A and late B stars 
is in preparation. 

In the region of the \ion{C}{1} lines, however, the synthetic spectrum
does fit the data well. 
The synthetic spectrum was shifted to the $-21$~\kms\ heliocentric 
velocity of the star, then aligned with the co-added LiF~1b spectra in 
the region where the photospheric \ion{C}{1} lines appear, using the 
cross-correlation procedure mentioned above. 
This process allowed us to establish the small zero-point offsets of 
the absolute wavelength calibration for the LiF~1 spectra with an 
uncertainty of no more than 14~\kms. 
Since the \oph\ spectrum has few photospheric absorption lines 
shortward of 1100~\AA, the offsets of the absolute wavelength calibration 
for the SiC~2b spectra were determined using the narrow IS/CS 
absorption lines of \ion{Ar}{1}. Thus, the uncertainty in the absolute 
calibration for SiC~2b is larger, about 27~\kms, due to the
spread in velocity between the unresolved IS and CS components.

Data taken with \fuse\ between about 2000 August and 31 July 2001,
when the detector high voltage was raised for the second time,
may suffer from an effect called detector x-walk. 
This effect causes the position
of a low pulse height photon event to be misplaced in the dispersion
direction, and was caused by excess gain sag in localized 
places on the detectors where the brightest airglow lines fall. 
The x-walk causes a distortion in the shape of an absorption 
line that conserves photons but not equivalent width, and varies with time. 
No good post-aquisition remedies for this effect exist at this time. 
Thus, one must be very careful when quantitatively analyzing any 
absorption lines in \fuse\ spectra taken before the detector high 
voltage was increased, particularly when looking for time-variability. 
To deal with this problem, we carefully examined the shape of each line
of interest in the spectra from two different detector channels, 
and determined which portions of the data suffered from the x-walk 
effect. We excluded the few portions of the data that showed 
significant x-walk from our analysis. 

\section{Overview of FUV Spectrum}\label{sec:over}

The stellar flux from 51~Oph in the \fuse\ wavelength range drops
off sharply below 1110 \AA, as can be seen in Figure~\ref{fig:over}.  
The most prominent features are the broad \ion{C}{1} photospheric 
absorption lines between 1110 \AA\ and 1160 \AA.
No emission lines, or any other indications of stellar activity, 
are observed in the data. 
The non-photospheric absorption features in the \oph\ \fuse\ spectrum 
are listed in Table~\ref{tab:lines}. 
Since the absorption features from the local interstellar medium 
(ISM) are only separated from the 
stellar heliocentric velocity by about 6 \kms\ \citep{craw97}, 
we are unable to resolve them in our spectrum. 
However, many lines in the spectrum arise from excited energy levels, and 
therefore are not usually seen in spectra of the ISM.  
These excited species indicate that there is warm CS gas present near 51~Oph.

\section{Metastable species}\label{sec:meta}

\begin{figure}
\begin{center}
\vspace*{-1.0in}
\epsfig{file=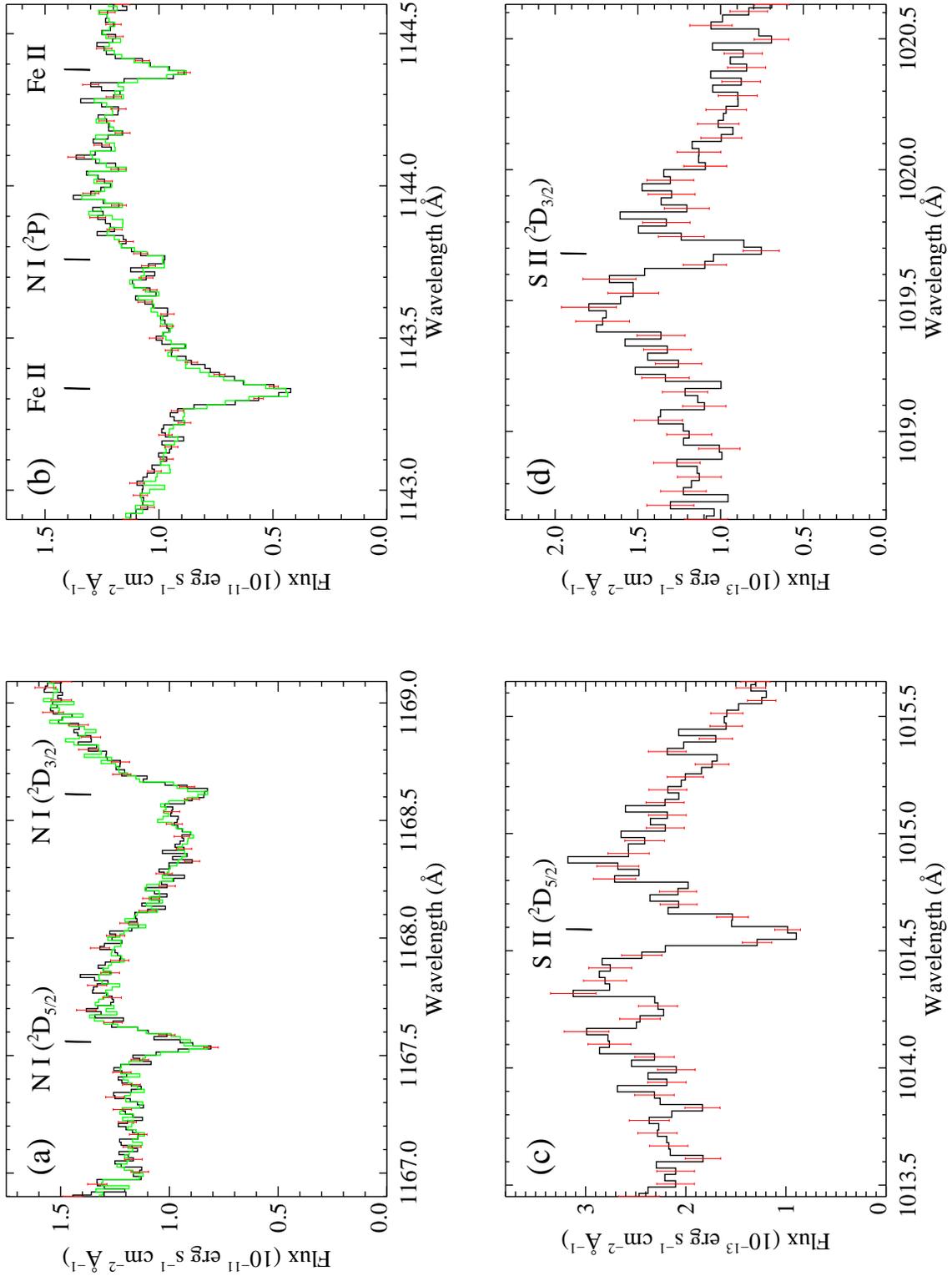, height=8.5in, angle=180}
\vspace*{0.6in}
\caption{Metastable species in the \oph\ CS gas. 
The $\pm \ 1 \sigma$ measurement error bars are 
overplotted in red. In the top two panels, data taken 2000~August~29 
are plotted in black and those taken 2000~September~4 in green.
None of the lines showed any significant variation between the two 
days of observation. In the bottom two panels, the data from the two days of 
observation were averaged together to improve the S/N ratio.
\ \ \textbf{(a)}~Two of the \ion{N}{1}~($^{2}D$) lines in the LiF~1b 
segment spectra, rebinned by a factor of 2. 
\ \ \textbf{(b)}~The blended \ion{N}{1}~($^{2}P$) \lb1143.65 doublet in the 
LiF~1b segment spectra, rebinned by a factor of 2. The \fuse\ spectral 
resolution is insufficient to separate the two lines of the doublet.
\ \ \textbf{(c)}~The \sdf\ \lb1014.44 line in the LiF~1a segment spectrum,
rebinned by a factor of 4. 
\ \ \textbf{(d)}~The \sdt\ \lb1019.53 line in the LiF~1a segment spectrum,
rebinned by a factor of 4. 
\label{fig:meta}}
\end{center}
\end{figure}

Several of the absorption lines arising from metastable states are shown
in Figure~\ref{fig:meta}. Only the strongest metastable line,
arising from \om, appeared to vary significantly between the two days 
of observation (shown in Figure~\ref{fig:FEBs}c).
For the other lines, we averaged the two days of data to improve 
the S/N ratio, and fit Gaussians to the lines to determine their redshifts.  
The strong \sd\ lines were slightly redshifted by $16 \pm 14$ \kms with
respect to the stellar velocity; the weaker \nd\ and \np\ lines 
appeared to be unshifted. We calculated the equivalent widths of the lines
from the data and converted them to column densities, assuming the
lines were unsaturated; these values appear in Table~\ref{tab:meta}.
For the \ndf\ lines, we had enough lines with different oscillator
strengths available to determine that the two weakest lines were 
unsaturated, so the column density for this species should be accurate. 
For the \ndt, \np, and \sd\ lines, this was not the case, so our
column densities for these species should be regarded as lower limits 
to the true values.
In particular, the \sd\ lines are probably considerably saturated.

The energies of the metastable states above the ground state are
2.38 eV for \nd, 3.58 eV for \np, and 1.84 eV for \sd. These
energies roughly correspond to temperatures of 28000 K, 41000 K, and
21000 K, respectively.
Therefore, these lines must arise solely from CS gas; additionally, this
gas is quite warm. We calculated the radiative decay lifetimes of the
metastable states by summing the probabilities of the spin-forbidden  
transitions to lower energy levels. 
These lifetimes appear in Table~\ref{tab:meta}. All the lifetimes are
quite a bit shorter than the six day interval between the two
observations; therefore, the metastable levels must be continuously
repopulated by collisional excitation, UV pumping, and/or 
molecular dissociation. Lines arising from the metastable $^1D$ level of
\ion{C}{1} are seen in spectra of \bp\ \citep{rob00} and solar system
comets \citep{toz98}. We intend to more fully investigate the significance
of metastable species in young CS disks in a future paper.

\section{Molecular gasses}\label{sec:mol}

No molecular hydrogen or carbon monoxide is observed toward the star.  
The $3 \sigma$ upper limits on the \hh\ line-of-sight column densities 
in the $J=0$ and $J=1$ rotational levels are 
${\rm N}(J=0) \leq 3 \times 10^{13}$ \cm\ and 
${\rm N}(J=1) \leq 4 \times 10^{13}$ \cm.  
The non-detection of absorption in the \mbox{C--X} (0,0) band at 
1088 \AA\ allows us to set an upper limit on the line-of-sight 
column density of $^{12}$CO in the ground 
vibrational level, ${\rm N}({\rm ^{12}CO}) \lesssim 7 \times 10^{13}$ \cm. 
These low upper limits for the line-of-sight column densities of the two most 
abundant molecular gasses are somewhat confusing in light of the 
recent detection of very large amounts of warm molecular gasses in {\it ISO} 
spectra of 51~Oph. Van den Ancker et al. (2001) detect emission 
from $^{12}$CO, $^{13}$CO, H$_2$O, CO$_2$ and, surprisingly, 
NO at a gas temperature of about 850 K. 
Our upper limit on ${\rm N}({\rm ^{12}CO})$ is 
highly inconsistent with their column density of $3 \times 10^{21}$ \cm. 
Even at 850 K, such a large column 
density of $^{12}$CO should produce a \mbox{C--X} (0,0) absorption band 
that is completely saturated over several Angstroms. 
This is clearly not seen in our data. Our non-detection of the
warm CO seen by {\it ISO} indicates that very little of it 
lies along the line of sight to the star through the edge-on disk.

Van den Ancker et al.\ (2001) propose two possible explanations for
the large amount of molecular gas they detected,
1) that a large mass ejection from the star into the disk 
has recently occurred, or
2) that a planet-sized body orbiting the star has recently 
been destroyed, possibly by a collision with another similar body.
Although the second scenario seems very exotic, it is more consistent with
our conclusion that very little of the warm CO seen by {\it ISO} 
lies along the line of sight. 
However, this issue must be thoroughly investigated with more sensitive 
observations searching for absorption from CO and other
molecular gasses. 

\section{Variable Features and Infalling Gasses}\label{sec:var}

\begin{figure}
\begin{center}
\vspace*{-0.2in}
\epsfig{file=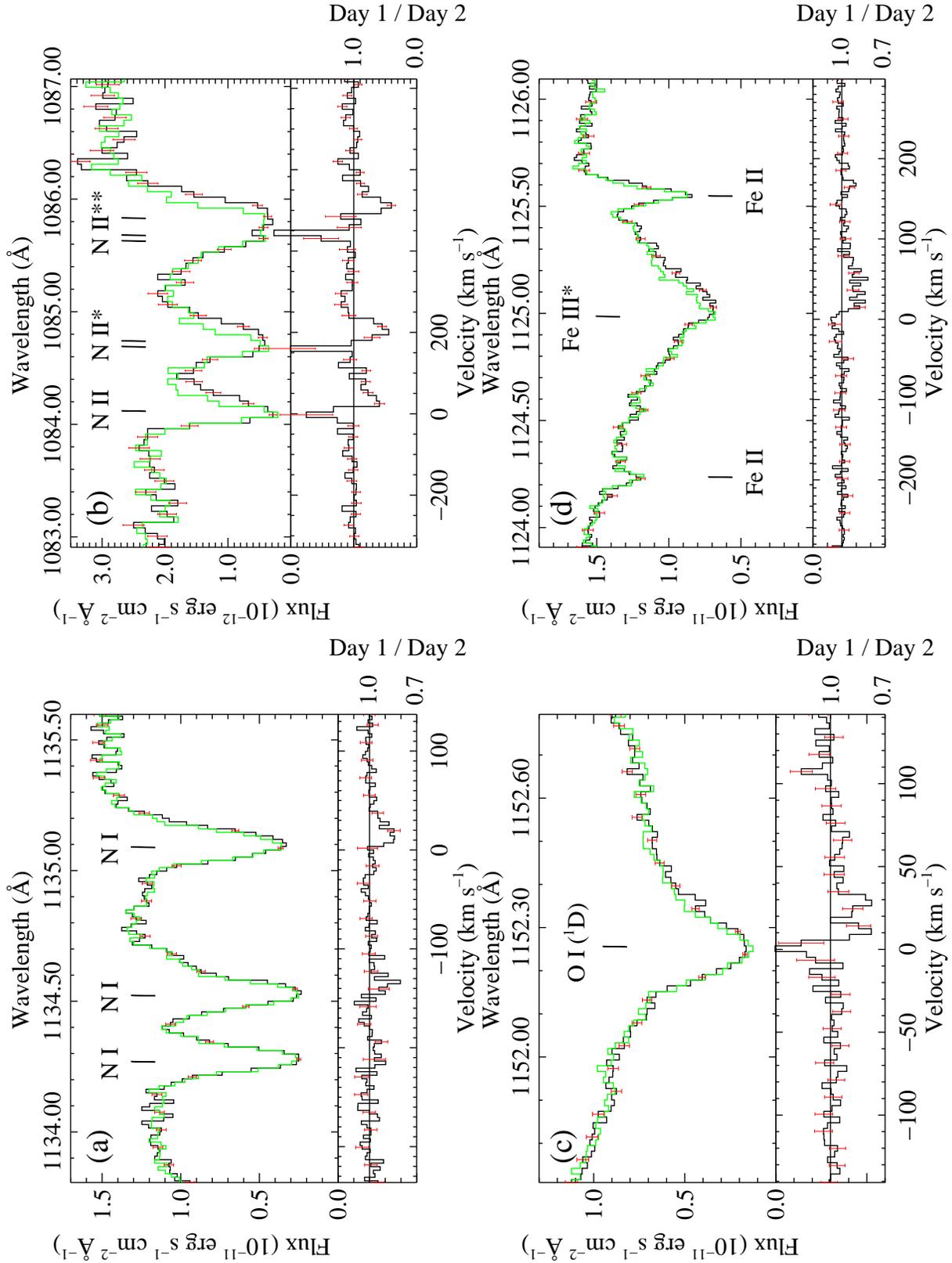, height=8.5in, angle=180}
\vspace*{0.6in}
\caption{Time-variable absorption lines in the 51~Oph \fuse\
spectra.\label{fig:FEBs} Data taken 2000~August~29 (black) 
and 2000~September~4 (green) are shown at the top of each panel.
The August 29 spectrum divided by the September 4 spectrum is shown at the 
bottom of each panel. The $\pm \ 1 \sigma$ measurement error bars are 
overplotted in red. Velocities are given in the rest frame of the star. 
\ \ \textbf{(a)}~The \ion{N}{1} $\lambda 1135$ triplet in the LiF~1b segment
spectra, rebinned by a factor of 2 for this plot. 
\ \ \textbf{(b)}~The \ion{N}{2} $\lambda 1085$ multiplet in the SiC~2b 
segment spectra, rebinned by a factor of 8.
\ \ \textbf{(c)}~The \ion{O}{1} ($^{1}D$) \lb1152.15 line in the
LiF~1b segment spectra, rebinned by a factor of 2.
\ \ \textbf{(d)}~The \ion{Fe}{3}* $\lambda 1124.87$ line in the LiF~1b segment
spectra, rebinned by a factor of 2.}
\end{center}
\end{figure}

\begin{figure}
\begin{center}
\hspace*{-0.2in}
\epsfig{file=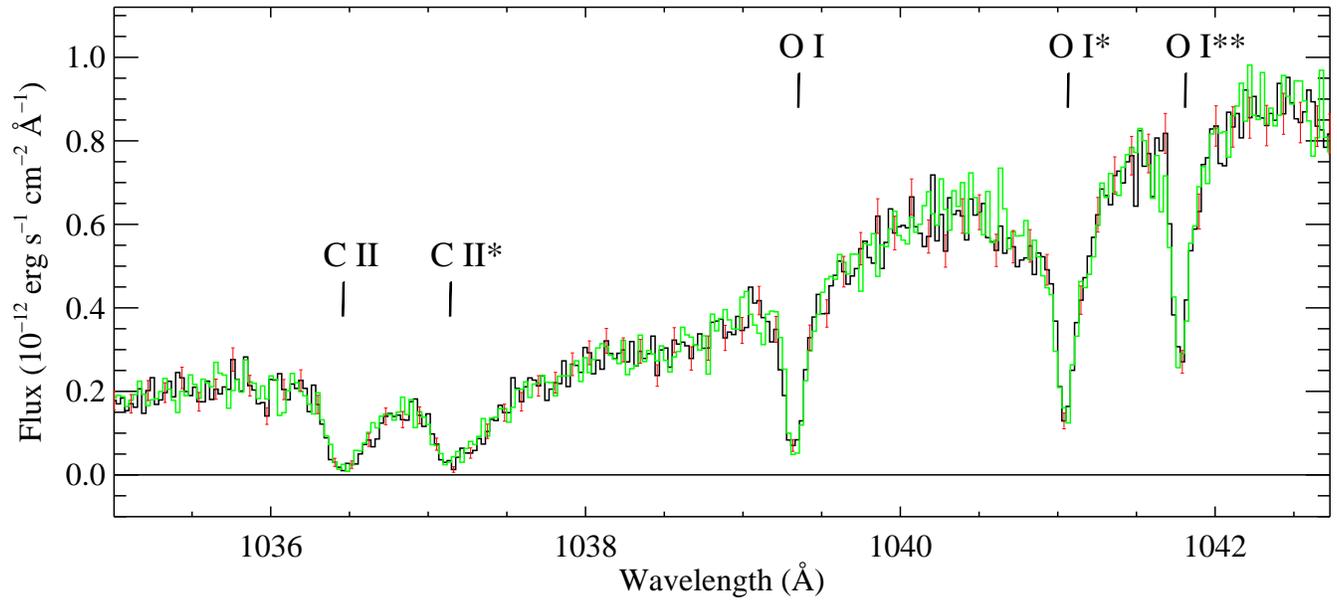}
\caption{Unvarying absorption lines in the \oph\ LiF~1a spectra, showing
data taken 2000~August~29 (black) and 2000~September~4 (green). No variable 
infalling gas is seen in these lines arising from carbon and oxygen. The data
were rebinned by a factor of 4 for this plot. The $\pm \ 1 \sigma$ 
measurement error bars are overplotted in red. \label{fig:non}}
\end{center}
\end{figure}

Comparison of the spectra from the two days of observation shows that
they are practically identical. However, careful examination of the high
S/N spectra revealed a few narrow absorption features that varied 
between the two days. They are shown in Figure~\ref{fig:FEBs}, 
the \ion{N}{1} $\lambda 1134$ triplet in Figure~\ref{fig:FEBs}a, 
the \nii\ $\lambda 1085$ multiplet in Figure~\ref{fig:FEBs}b, 
the line arising from metastable \om\ in Figure~\ref{fig:FEBs}c, 
and one of the four \feiii\ lines (1124.87 \AA) in Figure~\ref{fig:FEBs}d. 
No other lines in the \oph\ spectrum showed any significant time variation. 
For comparison, unvarying lines arising from \ion{C}{2} and
ground term \ion{O}{1}~($^3P$) are shown in Figure~\ref{fig:non}.
Note that most of the variable lines observed arise from excited 
energy levels, and therefore must represent purely CS gas.

In the August 29 spectrum, all the variable features show excess 
absorption that is redshifted with respect to the stellar velocity. 
The variable infalling gas absorption was isolated by dividing 
the August~29 (day~1) spectrum by the quiescent September~4 (day~2) 
spectrum, removing any possible IS, stellar, or non-varying CS 
absorption; the result is shown at the bottom of each panel of 
Figure~\ref{fig:FEBs}. The spectra from the two days were
cross-correlated using unvarying lines to align them exactly 
before normalization, but no shifting was necessary to align the LiF~1
spectra. All further analysis was performed on these normalized
spectra.

The apparent ``emission features'' in the \nii\ and \om\ normalized
spectra actually represent excess absorption in the day~2
spectrum, in a velocity range that is blueshifted relative to 
the day~1 excess absorption. Although they appear 
to be real because they are seen on the blue edge of all three \nii\
features, they are not resolved and are only marginally
(less than $3 \sigma$) detected.  No column densities were calculated
for these day~2 excess absorption features; data with higher spectral
resolution would have been necessary to confirm their reality.

The equivalent widths of the day~1 excess absorption features, calculated 
from the normalized spectrum, are shown in Table~\ref{tab:FEBs}. 
They were then converted into column densities, assuming that the 
lines were unsaturated, and using the revised absorption oscillator 
strengths of \citet{mor99}.  These column densities represent lower 
limits to the true column densities. 
More accurate measurement of the column densities was performed using 
the apparent optical depth method (see \citet{sav91} for a complete
discussion).   
The apparent optical depth, in velocity space, is
\begin{equation} \tau_a(v) = \ln \frac{I_o(v)}{I_{\rm obs}(v)} , \end{equation}
where $I_o(v)$ is the intensity without absorption, and $I_{\rm obs}(v)$
is the observed intensity. The apparent optical depth
will be a good representation of the true optical depth as long as no 
unresolved saturated structure is present in the absorption line. 
Unfortunately, for most of our lines we cannot
prove this directly, as we do not have two well separated lines differing
in $f \lambda$ arising from the same species and energy level. 
In our case, however, the excess absorption extends over 
at least 40 \kms, and is therefore likely resolved. So, the apparent 
column density in some particular velocity range from $v_1$ to $v_2$ is
\begin{equation} N_a = \int_{v_1}^{v_2} N_a(v) dv = 
\frac{m_{e} c}{\pi e^2 f \lambda} \int_{v_1}^{v_2} \tau_a(v) dv .
\end{equation} This method gives accurate column densities for lines that
are not strongly saturated.  A curve of growth analysis was
not possible, because we did not observe enough lines arising from the 
same species and energy level. Since the \nii* and \nii** lines are 
blended and may be strongly saturated, we were not able to perform 
a reliable apparent optical depth analysis on them. However, the equivalent 
width analysis at least provides lower limits on the column densities of 
\nii* and \nii**.

The velocity ranges of the day~1 excess absorption features and the 
apparent column densities in these ranges are shown in Table~\ref{tab:FEBs}.
First, we can see that many of the lines are unsaturated, because the
column densities obtained from the equivalent widths are equal
(within $1 \sigma$) to the apparent column densities. 
Second, the column densities of the \feiii\ lines show that the
energy levels are statistically populated, within measurement
errors. Therefore, the excitation temperature of the infalling \feiii\ 
gas must be greater than 1340 K, the temperature of the highest 
energy level observed. The presence of \om\ absorption implies
temperatures of around 23000 K. 

The total column densities of infalling \ion{N}{1} and \feiii\ gas, 
in all fine-structure levels, are 
$N_{\mathrm{N \; \scriptscriptstyle{I}}} = (3.06 \pm 0.47) 
\times 10^{13}$ \cm\ and $N_{\mathrm{\, Fe \; \scriptscriptstyle{III}}} 
= (2.66 \pm 0.21) \times 10^{14}$ \cm.
For \nii, we find the total column density in all fine-structure levels to 
be at least $(2.26 \pm 0.17) \times 10^{14}$ \cm. The maximum possible column 
density of \nii\ may be calculated by assuming a statistical population 
of the fine-structure levels (1:3:5); the total column 
density in all levels must be less than or equal to
$(6.7 \pm 1.0) \times 10^{14}$ \cm. We conclude that the total column 
density of infalling \nii\ gas, in all fine-structure levels, 
is in the range $(2.1 \: \sbond \: 7.7) \times 10^{14}$ \cm.

\section{Discussion}\label{sec:dis}

Previous workers have suggested that 51~Oph might have a companion
star, because its radial velocity was found to be variable 
\citep{bus63}. Our detection of time-variable, redshifted CS absorption 
lines in \fuse\ spectra of 51~Oph confirms the \citet{gra93} 
detection of such lines in {\it IUE} spectra. This confirmation
leads us to believe that any previously measured variation 
in the star's radial velocity is not due to the presence
of a stellar companion, but rather due to the appearance and
disappearance of redshifted CS absorption features. 

\subsection{Composition of the transient infalling gas}\label{sec:comp}

The ionization energies required to produce \feiii\ and \nii\ are 
16.19 eV and 14.53 eV, respectively. Since photons with 
wavelengths less than 766 \AA\ and 853 \AA\ are required to 
produce these species through photoionization, appreciable quantities 
cannot be produced by the IS UV field. 
Also, as can be seen from the \fuse\ spectrum of 51~Oph, the star
has virtually no flux below about 1040 \AA; therefore, these
species cannot be produced by stellar photoionization either.
They must be produced by collisional processes involving relatively dense 
gas. The fact that \nii\ is more abundant than \ion{N}{1}
suggests a moderately high temperature for the infalling gas. 
A detailed analysis of the ionization balance is beyond the scope
of this paper, but using the cooling functions of \citet{suth93} 
and assuming collisional ionization equilibrium, 
the ion fraction of \nii\ suggests a temperature between 
20000 and 34000 K for the transient infalling gas, which is consistent 
with the presence of \om\ in this gas.

The composition of the transient infalling gas appears to be 
rather unusual. No \ion{C}{1}, \ion{C}{2}, or \ion{C}{3} 
is detected in the transient infalling gas. 
In fact, the only carbon lines
observed in the \fuse\ \oph\ spectrum are photospheric \ion{C}{1} lines 
and \ion{C}{2} and \ion{C}{2}* ground term ($^{2}P$) lines arising 
from IS/CS gas, which do not vary significantly between the two days. 
The strong metastable \ion{C}{2}~($^{4}P$) line at 1010.37 \AA\ 
is not seen, nor are the strong \ion{C}{3} lines at 977.02 and 1175.71 \AA.
Given the estimated gas temperature range above, more than 80\% of the 
carbon should be present as \ion{C}{2} \citep{suth93}. 
We find a conservative $3 \sigma$ upper limit of $4 \times 10^{13}$~\cm\
for the column densities of both \ion{C}{2} and 
\ion{C}{2}* in the infalling gas, giving an upper limit on 
the total carbon column density, 
$N_{\rm{C}} \: \leq 1 \: \times 10^{14}$~\cm. 

The total nitrogen column density in the infalling gas, 
$N_{\rm{N}} = N_{\mathrm{N \; \scriptscriptstyle{I}}} 
+ N_{\mathrm{N \; \scriptscriptstyle{II}}}$, is in the range 
$(2.36 \: \sbond \: 8.05) \times 10^{14}$~\cm.
Therefore, the ratio of $N_{\rm{N}}$ to $N_{\rm{C}}$ in the 
transient infalling gas is at least 2.36.
The solar abundance of nitrogen relative to carbon,
$N_{\rm{N}}/N_{\rm{C}}$, is 0.251 \citep{grev90, grev91}. 
The mean gas phase abundance of carbon atoms in the Local ISM 
is $140 \pm 20$ per $10^{6}$ H atoms \citep{car96} and that of nitrogen is
$75 \pm 4$ per $10^{6}$ H atoms \citep{mey97}. These values do not
appear to vary with direction or with the physical conditions of the
IS environment. Therefore, the infalling gas in the 51~Oph system 
does not appear to be composed of unprocessed IS gas, nor does it 
have a solar-like composition. 

Given the gas temperature range determined from the nitrogen
ionization balance, 63\% to 64\% of the total iron should
be present as \feiii\ \citep{suth93}. 
The total column density of iron in the infalling gas, $N_{\rm{Fe}}$,
is therefore roughly $(3.83 \: \sbond \: 4.56) \times 10^{14}$ \cm.
The ratio of $N_{\rm{Fe}}$ to $N_{\rm{C}}$ in the 
transient infalling gas is at least 3.83, while the solar
abundance of iron relative to carbon is 0.0795 \citep{grev99},
indicating that iron is even more overabundant relative to carbon than 
nitrogen is. 
We consolidate the results of our abundance analysis below.
\vspace{8pt}
\[ \frac{ (N_{\rm{N}}/N_{\rm{C}})_{\rm{\oph}}}
{(N_{\rm{N}}/N_{\rm{C}})_{\rm{solar}}} \geq 9.4 \]
\vspace{8pt}
\[ \frac{ (N_{\rm{Fe}}/N_{\rm{C}})_{\rm{\oph}}}
{(N_{\rm{Fe}}/N_{\rm{C}})_{\rm{solar}}} \geq  48.2 \]
\vspace{8pt}
\[ \frac{ (N_{\rm{Fe}}/N_{\rm{N}})_{\rm{\oph}}}
{(N_{\rm{Fe}}/N_{\rm{N}})_{\rm{solar}}} = (1.5 \: \sbond \: 6.1) \]
We were not able to include the oxygen
abundance in this analysis, because at temperatures of 20000 to
34000 K, less than 8\% of the oxygen is present as \ion{O}{1}.
Therefore, we require a large extrapolation from our measured 
\om\ column density to obtain the total oxygen column density. 
Unfortunately, there are no ground term \ion{O}{2} lines longward 
of 911.7 \AA.

\citet{nat00} establish a useful test to determine if gas infall
is related to planetesimal evaporation or magnetospheric
accretion from a massive CS disk, as seen in classical T Tauri
stars \citep{hart94}. 
If the infalling gas is produced from star-grazing planetesimals, it
should be depleted in hydrogen (or alternatively, it should 
be metal-rich). On the other hand, one would expect a roughly 
solar composition if the redshifted absorption is due to
magnetospheric accretion.
While our analysis of the composition of the infalling gas seen with 
\fuse\ is somewhat incomplete since we have no information on the 
hydrogen abundance, it does indicate that the composition of the 
transient infalling gas is extremely non-solar. 
Specifically, it appears that the gas must be either volatile-depleted or
iron-rich, indicating that the gas is more likely to have been produced by
vaporization of planetary material.

\subsection{Dynamics of the transient infalling gas}\label{sec:dyn}

We calculated the ratio of the force of radiation pressure from
51~Oph to the force of gravity for each variable species observed,
using the equations 
\begin{equation} F_{rad}  \; = \sum_{all \ transitions} \frac{1}{4 \pi 
\epsilon_0} \frac{\pi e^{2}}{m_{e} c^{2}} f_{i} \Phi_{\lambda_{i}}, 
\end{equation}
\begin{equation} F_{grav} = \frac{G M_{star} M_{atom}}{r^2}, \hspace{10pt}
\beta = \frac{F_{rad}}{F_{grav}}, \end{equation}
where $f_i$ is the absorption oscillator strength, $\Phi_{\lambda_{i}}$ 
is the stellar flux at some distance $r$ from the star, $M_{star}$ is 
the mass of 51~Oph, and $M_{atom}$ is the mass of the atomic species. 
Note that the ratio of the forces, $\beta$, is independent of 
the distance $r$.
The stellar fluxes were either measured from the \fuse\ spectrum,
or estimated from our synthetic stellar spectrum at longer wavelengths.
The strongest transitions are in the \fuse\ wavelength range, and
the values of $\beta$ do not depend sensitively on the synthetic spectrum 
fluxes.
We used a stellar mass of $3 \ \rm{M}_\sun$, estimated from the star's
position on the H-R diagram and assuming that it is on the
zero-age main sequence. The calculated values of $\beta$ are shown in 
Table~\ref{tab:FEBs}.
We can see that for most of the variable species
$\beta$ is greater than 1, and for \ion{N}{1} and \om\ it is quite large.
One would therefore generally expect any optically-thin absorption from 
these species to be blueshifted, unless the gas was injected toward the star 
at a relatively high radial velocity, as would happen if it was
produced from infalling planetesimals.  

\section{Concluding Remarks}\label{sec:end}

A comparison of the characteristics of 51~Oph 
to those of $\beta$~Pictoris shows many similarities, 
strengthening the suggested evolutionary connection between the two 
systems. Both CS disks contain highly excited, metastable species
that have short radiative decay lifetimes.
Both systems also show the presence of variable, collisionally ionized 
CS material. The variable absorption features in both systems
are most frequently redshifted with respect to the stellar velocity, 
despite the effect of radiation pressure. There are some differences 
between the two systems, however.  If the variable redshifted events in 
the 51~Oph system are caused by infalling planetesimals, this system has
a lower infall rate than $\beta$~Pic, since many observations of 
51~Oph have not shown any variable infalling gas features 
\citep{lecav97b, craw97}. 

The composition of the transient infalling gas in \oph\ is 
highly non-solar, and is more likely produced by the destruction 
of a star-grazing planetesimal than by magnetospheric accretion of 
unprocessed gas.  However, this conclusion needs to be tested by 
observation of additional infalling gas events in lines arising from
many species, in order to perform a more complete analysis of the 
infalling gas composition. 
And detailed modeling is necessary to fully 
understand the significance of the metastable atomic species in both
\bp\ and \oph. Identification of \oph\ as a planetary, rather than 
protoplanetary, system will at last firmly add another object to the 
class of \bp-type systems. 
Since \oph\ is much younger than \bp, which is about 
$2 \times 10^{7}$ yrs old \citep{bar99}, this will provide a tighter 
constraint on the time scale for planetary formation, and increase 
our knowledge about the evolution of CS disks.

\acknowledgements
We thank Pierre Chayer for providing us with a rotationally 
broadened synthetic stellar spectrum for 51~Oph. We also thank Chris Howk for 
helpful discussions about applying the apparent optical depth method
to our dataset. This work is based on observations made with the
NASA/CNES/CSA {\it Far Ultraviolet Spectroscopic Explorer}, operated
for NASA by the Johns Hopkins University under NASA contract
NAS5-32985.

\clearpage

\begin{deluxetable}{lccc}
\tablecolumns{4}
\tabletypesize{\small}
\tablewidth{0pt}
\tablecaption{IS and CS absorption lines in the \fuse\ spectrum of 
51~Oph\label{tab:lines}}
\tablehead{\colhead{Species} & \colhead{Wavelength} & \colhead{Energy Level} & 
\colhead{Comments} \\ 
\colhead{} & \colhead{(\AA)} & \colhead{(cm$^{-1}$)} & \colhead{}}
\startdata
\ion{C}{2} & 1036.34 & 0.00 & part IS \\
\ion{C}{2}* & 1037.02 & 63.42 & purely CS \\
\ion{O}{1} & 1039.23 & 0.00 & part IS \\
\ion{O}{1}* & 1040.94 & 158.27 & purely CS \\
\ion{O}{1}** & 1041.69 & 226.98 & " \\
\ion{Si}{2} & 989.87, 1020.70 & 0.00 & part IS \\
\ion{Si}{2}* & 992.69, 1023.70 & 287.24 & purely CS \\
\ion{Ar}{1} & 1048.22, 1066.66 & 0.00 & part IS \\
\ion{Fe}{2} & 1142.37, ..., 1055.26 & 0.00 & part IS \\
\ion{Fe}{2}* & 1146.95, ..., 1060.44 & 384.79 & purely CS \\
\ion{Fe}{2}** & 1150.29, ..., 1063.02 & 667.68 & " \\
\ion{Fe}{2}*** & 1153.27, ..., 1065.84 & 862.61 & " \\  
\ion{Fe}{2}**** & 1154.40, ..., 1076.85 & 977.05 & " \\
\sidehead{{\it Metastable Species}}
\ion{N}{1} \ $(^{2}D_{5/2})$ & 1163.88, ..., 1176.51 & 19224.47 & purely CS \\
\ion{N}{1} \ $(^{2}D_{3/2})$ & 1168.54, 1177.70 & 19233.18 & " \\
\ion{N}{1} \ $(^{2}P)$ & 1143.646, 1143.651 & 28839 & " \\
\ion{S}{2} \ $(^{2}D_{3/2})$ & 1019.53 & 14852.87 & purely CS  \\
\ion{S}{2} \ $(^{2}D_{5/2})$ & 1014.44 & 14884.67 & " \\
\sidehead{{\it Time-Variable Species}}
\ion{N}{1} & 1134.17, 1134.41, 1134.98 & 0.00 & part IS \\
\ion{N}{2} & 1083.99 & 0.00 &  \\
\ion{N}{2}* & 1084.57, 1084.58 & 48.67 & purely CS \\
\ion{N}{2}** & 1085.53, 1085.55, 1085.71 & 130.80 & " \\
\ion{O}{1} \ $(^{1}D)$ & 1152.15 & 15867.86 &  purely CS, metastable \\
\ion{Fe}{3}  & 1122.52 & 0.00 &  \\
\ion{Fe}{3}* & 1124.87 & 435.80 & purely CS \\
\ion{Fe}{3}** & 1126.72 & 738.55 &  " \\
\ion{Fe}{3}*** & 1129.19 & 932.06 & " \\
\enddata 
\end{deluxetable}

\clearpage

\begin{deluxetable}{lcccccc}
\tablecolumns{7}
\tabletypesize{\small}
\tablewidth{0pt}
\tablecaption{Metastable CS species\label{tab:meta}}
\tablehead{\colhead{Species} & \colhead{Wavelength} & \colhead{Energy level} & 
\colhead{Lifetime} & \colhead{$W_{\lambda}$\tablenotemark{a}}
& \colhead{$N_{W_{\lambda}}$\tablenotemark{b}} & \colhead{Comments} \\
\colhead{} & \colhead{(\AA)} & \colhead{(cm$^{-1}$)} & \colhead{} &
\colhead{(m\AA)} & \colhead{(\cm)} & \colhead{} }
\startdata
\ndf & 1168.42 & 19224.47 & 48.2 hrs & $2.5 \pm 1.1$ 
& $(1.06 \pm 0.45) \times 10^{14}$ & unshifted \\
\ndt & 1177.70 & 19233.18 & 12.3 hrs & $8.4 \pm 1.3$ 
& $(5.07 \pm 0.78) \times 10^{13}$ & unshifted \\
\np & 1143.646, 1143.651 & 28839 & 13 sec & $7.98 \pm 0.92$ 
& $(5.77 \pm 0.66) \times 10^{12}$
& unshifted, blended\tablenotemark{a} \\
\sdt & 1019.53 & 14852.87 & 39 min & $54.0 \pm 7.6$ 
& $(1.15 \pm 0.16) \times 10^{15}$ & shifted\tablenotemark{b} \\
\sdf & 1014.44 & 14884.67 & 36 min & $86.7 \pm 5.8$ 
& $(1.68 \pm 0.11) \times 10^{15}$ & shifted\tablenotemark{b} \\
\enddata
\tablenotetext{a}{For the blended lines, we used an effective oscillator
strength equal to the sum of the two individual line oscillator strengths.}
\tablenotetext{b}{The metastable \ion{S}{2} lines are slightly redshifted by
$16 \pm 14$ \kms\ with respect to the stellar velocity.}
\end{deluxetable}


\begin{deluxetable}{lcccccccc}
\tablecolumns{9}
\tabletypesize{\small}
\tablewidth{0pt}
\tablecaption{Variable infalling gasses\label{tab:FEBs}}
\tablehead{\colhead{Species} & \colhead{Wavelength} & \colhead{Energy level} & 
\colhead{$W_{\lambda}$\tablenotemark{a}}
& \colhead{$N_{W_{\lambda}}$\tablenotemark{b}} 
& \colhead{Velocity range\tablenotemark{c}} 
& \colhead{$\beta$\tablenotemark{d}} 
& \colhead{$N_{a}$\tablenotemark{e}} & \colhead{Comments}  \\ 
\colhead{} & \colhead{(\AA)} & \colhead{(cm$^{-1}$)} & \colhead{(m\AA)} &
\colhead{($10^{13}$ \cm)} & \colhead{(\kms)} & \colhead{} & 
\colhead{($10^{13}$ \cm)} & \colhead{} }
\startdata
\ion{N}{1} & 1134.41  & 0.00 & $9.1 \pm 2.8$ & $2.68 \pm 0.83$ & 0 -- 40 
& 7.4 & $3.07 \pm 0.76$  & unsaturated \\ 
\ion{N}{1} & 1134.98  & " & $15.7 \pm 3.2$ & $3.16 \pm 0.65$ & " 
& " & $3.06 \pm 0.47$  & " \\ [0.08in]
\ion{N}{2} & 1083.99 & 0.00 & $58 \pm 13$  & $4.8 \pm 1.1$  & 14 -- 114 
& 1.6 & $7.4 \pm 1.2$  & saturated  \\
\ion{N}{2}* & 1084.57, 1084.58 & 48.67 & $85 \pm 11$ & $7.09 \pm 0.91$ 
& \nodata & 1.6 & \nodata  & blended\tablenotemark{f} \\
\ion{N}{2}** & 1085.53, 1085.55, 1085.71 & 130.80 & $95 \pm 11$ 
& $8.02 \pm 0.92$ & \nodata & 1.6 & \nodata & 
blended\tablenotemark{f} \\ [0.08in]
\ion{O}{1} \ $(^{1}D)$ \ \tablenotemark{g} & 1152.15 & 15867.86 
& $22.7 \pm 3.5$ & $1.84 \pm 0.28$ & 6 -- 71 & 4.7 & $1.85 \pm 0.30$ 
& unsaturated \\ [0.08in]
\ion{Fe}{3} & 1122.52  & 0.00 & $48.5 \pm 4.4$ & $7.99 \pm 0.73$ 
& 0 -- 84 & 0.25 & $9.42 \pm 0.85$ & saturated \\
\ion{Fe}{3}* & 1124.87 & 435.80  & $26.4 \pm 2.7$ & $6.58 \pm 0.67$
& " & 1.3 & $6.26 \pm 0.68$ & unsaturated \\
\ion{Fe}{3}** & 1126.72 & 738.55 & $11.4 \pm 3.1$ & $5.4 \pm 1.5$
& " & 0.68 & $6.6 \pm 1.5$  & " \\
\ion{Fe}{3}*** & 1129.19 & 932.06 & $16.9 \pm 4.3$ & $3.75 \pm 0.96$
& " & 0.46 & $4.32 \pm 0.97$  & "  \\ 
\enddata  
\tablenotetext{a}{Equivalent width.}
\tablenotetext{b}{Column density calculated from $W_{\lambda}$, 
assuming the line is unsaturated. These values are lower limits
to the true column densities.}
\tablenotetext{c}{Velocities given with respect to the stellar velocity.}
\tablenotetext{d}{Ratio of radiation pressure to gravity for this
species.}
\tablenotetext{e}{Apparent column density, obtained by integrating the 
apparent optical depth over the velocity range.}
\tablenotetext{f}{For the blended lines, we used an effective oscillator
strength equal to the sum of the individual line oscillator strengths.}
\tablenotetext{g}{Metastable species.}
\end{deluxetable}

\clearpage

\clearpage

\clearpage

\clearpage

\end{document}